\documentclass{jkas}

\def\year{2024} % publication year
\def\volume{??} % journal volume
\def\issue{?} % journal issue
\def\beginpage{1} % first page of article
\def\received{---} % date paper was received by JKAS
\def\accepted{---} % date of acceptance
\def\published{---} % date of publication
\date{Received \received; Accepted \accepted; Published \published}

\graphicspath{{./}{Figures/}}

\title{%
HCN and HNC in the Disk of an Outbursting Young Star, V883 Ori
}

\author[1]{Seonjae Lee}{0000-0001-6324-8482}
\author[1, 2,$\star$]{Jeong-Eun Lee}{0000-0003-3119-2087}
\author[3]{Seokho Lee}{0000-0002-0226-9295}
\affil[1]{Department of Physics and Astronomy, Seoul National University, 1 Gwanak-ro, Gwanak-gu, Seoul 08826, Korea}
\affil[2]{SNU Astronomy Research Center, Seoul National University, 1 Gwanak-ro, Gwanak-gu, Seoul 08826, Republic of Korea}
\affil[3]{Korea Astronomy and Space Science Institute, 776 Daedeok-daero, Yuseong, Daejeon 34055, Korea}

\def\corrauthor{%
Jeong-Eun Lee, \email{lee.jeongeun@snu.ac.kr}
}

\def\runningauthor{%
Lee et al.
}

\def\runningtitle{%
HCN/HNC of V883 Ori
}

\def\keywords{%
astrochemistry --- ISM: molecules  --- stars: formation
}

\def\abstracttext{%
Hydrogen cyanide (HCN) and hydrogen isocyanide (HNC) are isomers with similar chemical properties. However, HNC can be converted into other molecules by reactions with atomic hydrogen (H) and atomic oxygen (O), resulting in a variation of the HCN/HNC abundance ratio. These reaction rates are sensitive to gas temperature, resulting in different abundance ratios in different temperature environments. 
The emission of HCN and HNC was found to distribute along ring structures in the protoplanetary disk of V883 Ori. HCN exhibits a multi-ring structure consisting of inner and outer rings. The outer ring represents a genuine chemical structure, whereas the inner ring appears to display such characteristics due to the high dust continuum optical depth at the center. However, HNC is entirely depleted in the warmer inner ring, while its line intensity is similar to that of HCN in the colder outer ring. In this study, we present a chemical calculation that reproduces the observed HCN/HNC abundance ratio in the inner and outer rings. This calculation suggests that the distinct emission distribution between HCN and HNC results from a currently ongoing outburst in V883 Ori. The sublimation of HCN and HNC from grain surfaces and the conversion of HNC to HCN determine their chemical distribution in the heated, warm inner disk.
}

\begin{document}
\jkashead %% set title, authors, abstract, etc.

%%%%%%%%%%%%%%%%%%%%%%%%%%%%%%%%%%%%%%%%%%%%%%%%%%%%%%%%%%%%%%%%%%%%%
%%% BEGIN MAIN TEXT HERE %%%%%%%%%%%%%%%%%%%%%%%%%%%%%%%%%%%%%%%%%%%%
%%%%%%%%%%%%%%%%%%%%%%%%%%%%%%%%%%%%%%%%%%%%%%%%%%%%%%%%%%%%%%%%%%%%%

\section{Introduction} \label{sec:intro}
Hydrogen cyanide (HCN) and its isomer hydrogen isocyanide (HNC) are N-bearing simple molecules commonly found in the interstellar medium and star-forming regions. They serve as high-density tracers on a wide range of scales, from external galaxies \citep{Harada2024} to protoplanetary disks \citep{jelee2024}.

The formation of HCN and HNC is mainly due to the dissociative recombination of HCNH$^{+}$, with a branching ratio of 1:1 \citep{Mendes2012}. However, variations in the HCN/HNC abundance ratio have been observed in many different environments, including high-mass star-forming regions \citep{Jin2015} and the integral shape filament (ISF) in Orion \citep{Hacar2020}. 
It was also found on a protoplanetary disk, V883 Ori \citep{jelee2024}. Since the formation process of these two molecules is predicted to be similar, the destruction mechanisms control the variation in their abundance ratio. Two temperature-sensitive neutral-neutral reactions have been regarded as the primary causes of the abundance variation:
\begin{align}
    \mathrm{HNC + H} &\rightarrow \mathrm{HCN + H} \label{reacH} \tag{R1}\\
    \mathrm{HNC + O} &\rightarrow \mathrm{NH + CO} \label{reacO} \tag{R2}
\end{align}
These reactions either destroy HNC or convert HNC into HCN, thereby increasing the abundance ratio. \par
The barrier energies of these reactions are still a matter of debate. Quantum chemical calculations have been conducted to investigate these reactions. A study of reaction \ref{reacH} by \cite{Talbi1996} determined the barrier energy to be 2000\,K.

In a subsequent study, \cite{Graninger2014} proposed an updated barrier of 1200\,K. The reaction \ref{reacO} was examined by \cite{Lin1992}, who calculated the barrier energy to be approximately 1100\,K. However, the barrier energies obtained from observational data show different results. \cite{Schilke1992} proposed a barrier energy of 200\,K for reaction \ref{reacH}, and \cite{Hacar2020} suggested 20\,K for reaction \ref{reacO} in order to fit the observations. The reason for this discrepancy is still under investigation. KIDA\citep{Wakelam2012}, an astrochemical network used in many studies, uses a barrier energy of 2000\,K for reaction \ref{reacH} and 1130\,K for reaction \ref{reacO}. \par
The abundance ratio is currently regarded as a tracer for investigating gas properties. \cite{Hacar2020} suggested using the HCN/HNC ratio as a 'thermometer', based on their observation of the ISF in Orion. They propose a two-part linear relation between the kinetic temperature of the gas derived from observations of ammonia and the HCN/HNC ratio. 
Alternatively, \cite{Santamaria2023} and \cite{Harada2024} propose that the HCN/HNC ratio is a tracer of FUV, rather than temperature. \cite{Behrens2022} suggests that the cosmic ray ionization rate may positively correlate with the abundance ratio. Further modeling is required to verify these proposals, given the difference in the physical conditions of the observed regions. \par
\cite{Long2021} constructed chemical models for model disks of typical T Tauri and Herbig Ae stars. The simulated HCN and HNC emission exhibited a double ring structure, with the inner ring originating from the inner disk atmosphere and the outer ring from the outer disk midplane. However,  previously existing observations of protoplanetary disks (e.g., TW Hya and HD163296; \citet{Graninger2015} and Lupus; \citet{Tazzari2021}) were not sensitive enough to resolve ring structures of HCN and HNC.
A double ring structure has been only recently observed in the disk of the Herbig Ae star HD100546 \citep{Booth2024}. The ring structures coincided with the dust rings observed in the continuum. \cite{Leemker2024} attempted to construct a chemical model reproducing the double-ring structure but was unsuccessful; although manipulating the C/O ratios in certain regions could roughly reproduce the HCN rings, the model fail to reproduce the emission distributions of other molecules. These results demonstrate that modeling the structures of HCN and HNC is a challenging task, due to the interplay of theoretical and observational considerations.

Therefore, the objective of this paper is to reproduce the observed HCN/HNC abundance ratio in V883 Ori using a chemical model based on the representative physical conditions of V883 Ori, instead of constructing a self-consistent two-dimensional (2D) physical and chemical model.  This allows us to isolate the essential chemical reactions that form the observed HCN and HNC emission structures. Furthermore, we investigate the effects of external physical parameters (UV strength and cosmic ray ionization rate) on the distributions of HCN and HNC observed in V883 Ori. 

\section{Previous observations of V883 Ori} \label{sec:v883}

\begin{figure*}[h]
    \centering
    \includegraphics[width=0.4\textwidth]{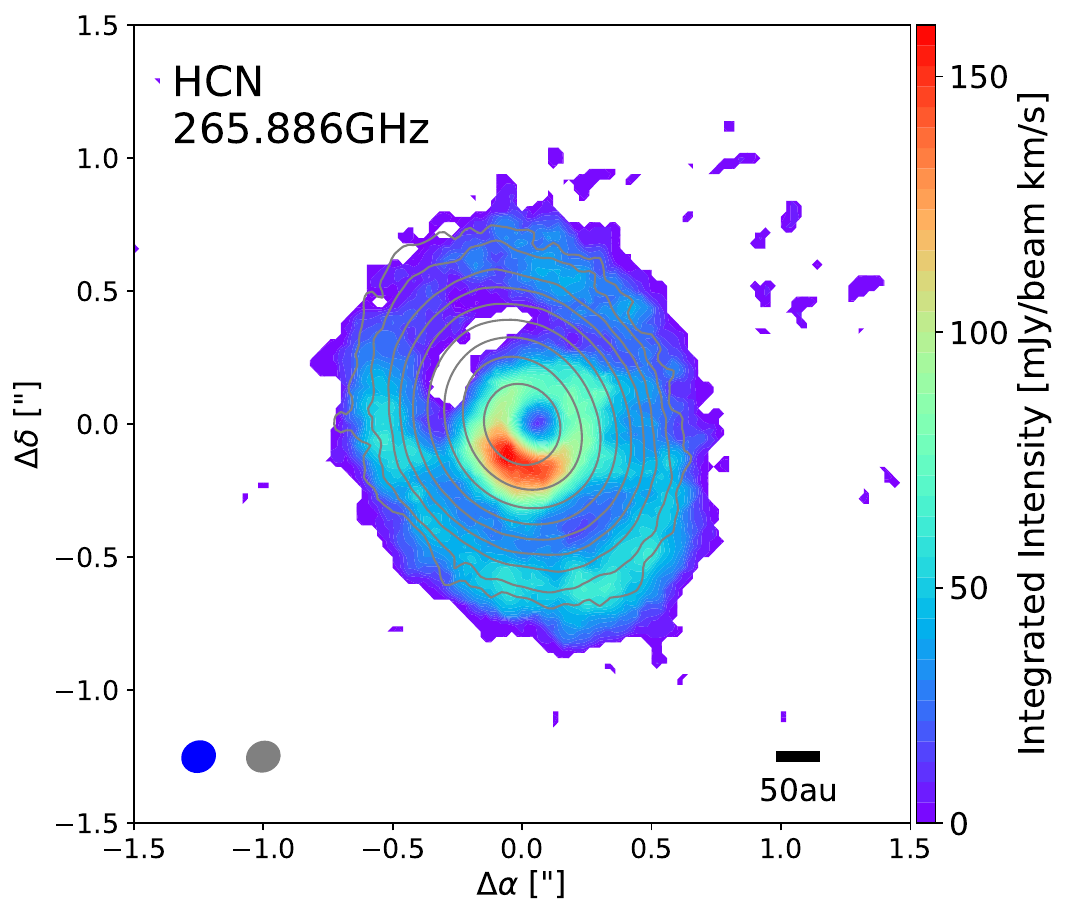}
    \includegraphics[width=0.4\textwidth]{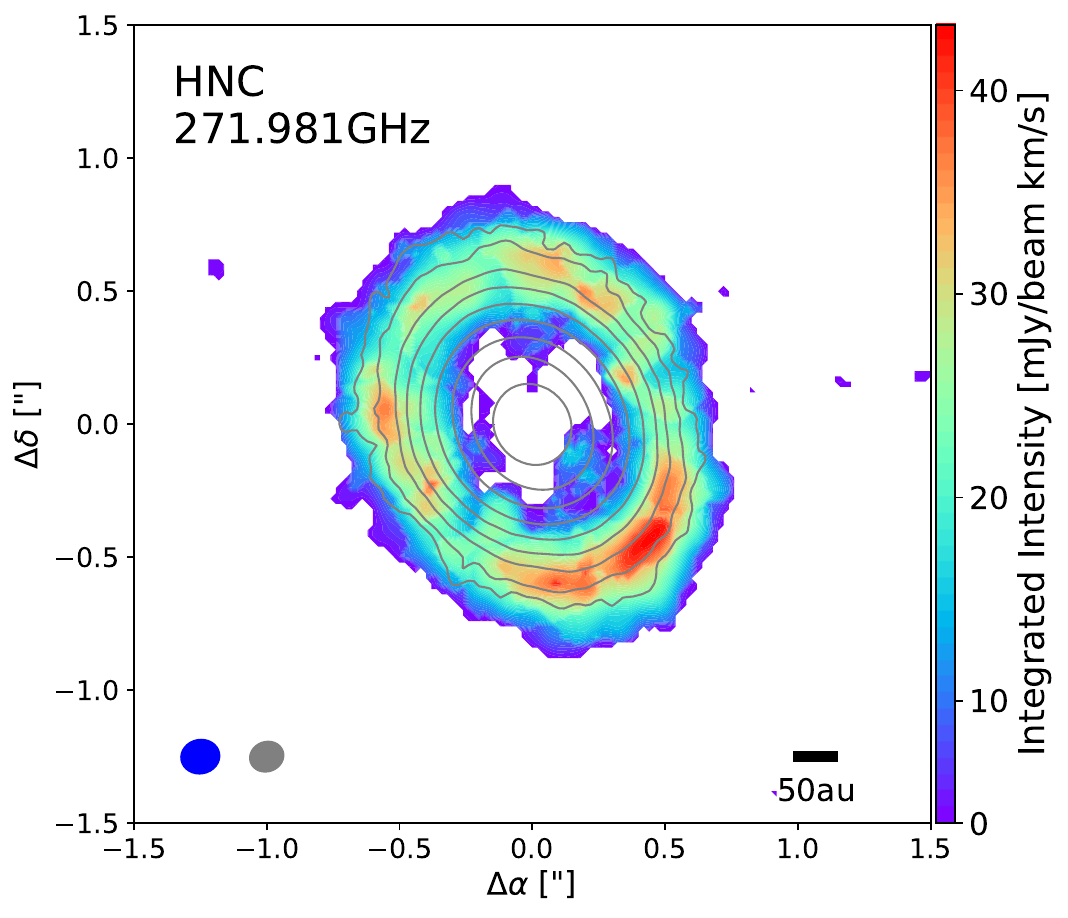}
    \includegraphics[width=0.4\textwidth]{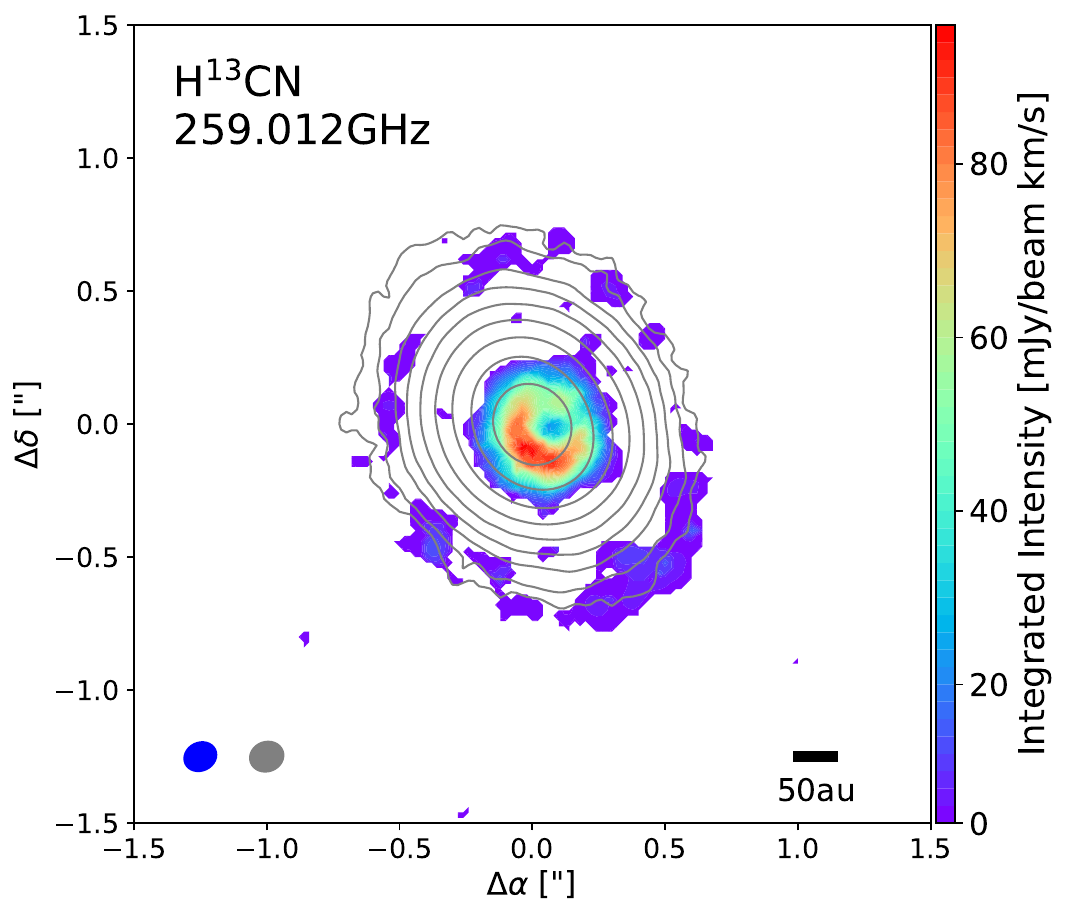}
    \includegraphics[width=0.4\textwidth]{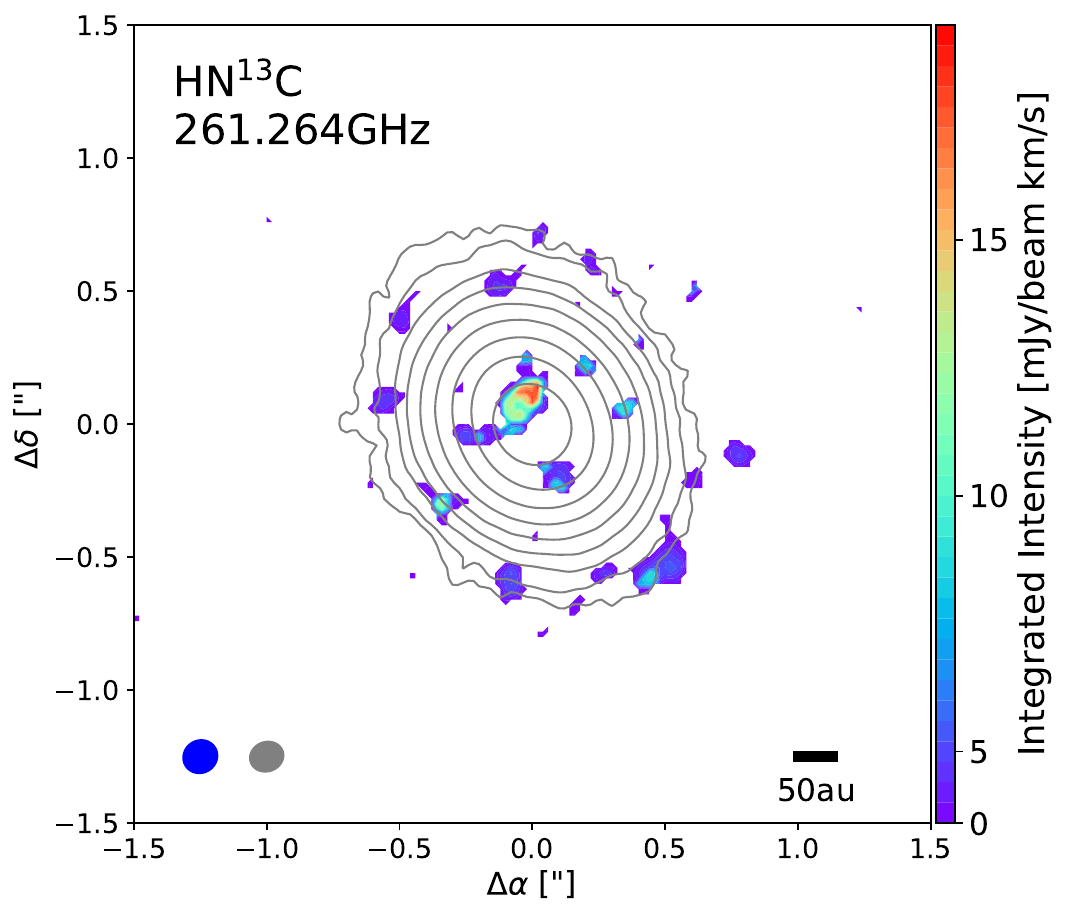}
    \caption{(Top) Detected HCN (left) and HNC (right) emission integrated intensity (moment 0) maps of V883 Ori \citep{jelee2024}. The line emission intensity below 3\,$\sigma$ is clipped out. The emission is integrated within $\pm 3$\,km\,s$^{-1}$ of the source velocity (4.3\,km\,s$^{-1}$). The continuum is overlayed in gray lines. The contour levels are 5, 10, 25, 50, 100, 200, 400, 800, and 1600\,$\sigma$ ($\sigma$ = 0.023\,mJy\,beam$^{-1}$). The beam size for the emission line and continuum are depicted in the lower left corner in blue and gray ellipses, respectively. Note that the color bar is different, so the HCN emission is generally stronger than HNC. (Bottom) It is the same with the top panels but with their $^{13}$C isotopologues. At the moment 0 map of HN$^{13}$C in the lower right, the emission inside the water sublimation region originates from CH$_2$DOH, whose line is located at the same spectral range of the HN$^{13}$C line. }
    \label{fig:V883}
\end{figure*}

V883 Ori is a sun-like \citep[M=\,1.2\,M$_\odot$,][]{Cieza2016, jelee2019} young stellar object in between Class I and II. 
It is currently going through an accretion burst.
The bolometric luminosity of V883 Ori is controversial, with estimates ranging from 186\,L$_\odot$ \citep{Furlan2016} to 285\,L$_\odot$ \citep{Strom1993, Sandell2001} when they are adjusted to a distance of 388\,pc \citep{jelee2019}. 
It is the first disk source where complex organic molecules (COMs) other than methanol (CH$_3$OH) and methyl cyanide (CH$_3$CN) have been detected in the midplane \citep{vantHoff_2018,jelee2019}. This object was observed by the ASSAY (ALMA Spectral Survey of An eruptive Young star, V883 Ori) project on ALMA Band 6 in Cycle 7 (2019.1.00377.S, PI: Jeong-Eun Lee) \citep{jelee2024}. The observation of the spectral windows containing HCN and HNC lines was performed on November 15, 2021. All spectral windows had spectral resolution of 488.281\,kHz ($\sim$0.55\,km\,s$^{-1}$). We used natural weighting to clean the images for a better resolution. Both line data had a similar beam size of $\sim$0.14"$\times$0.12", and the rms noise level of $\sim$2.2\,mJy\,beam$^{-1}$. Further details regarding the observation and data reduction can be found in \citet{jelee2024}.

The top panels of Figure \ref{fig:V883} depict the integrated intensity (moment 0) maps of HCN J=3-2 (265.886\,GHz) and HNC J=3-2 (271.981\,GHz). The HCN line has hyperfine structures, and all hyperfine lines are integrated. The HCN emission exhibits a double ring structure, with the outer ring located at a radius of $\sim$0.65" and the inner ring positioned at $\sim$0.2" along the semi-major axis. It should be noted that the inner hole within $\sim$0.1" is due to the optically thick dust continuum emission, as analyzed in previous studies \citep{jelee2019, jelee2024, RuizRodriguez2022}. HNC does not have an inner ring but has an outer ring in a similar location (r$\sim$0.6") with HCN. The position difference between the peak emission of HCN and HNC in the outer ring is much smaller than the beam.\par

\begin{figure}[h]
    \centering
    \includegraphics[width=\columnwidth]{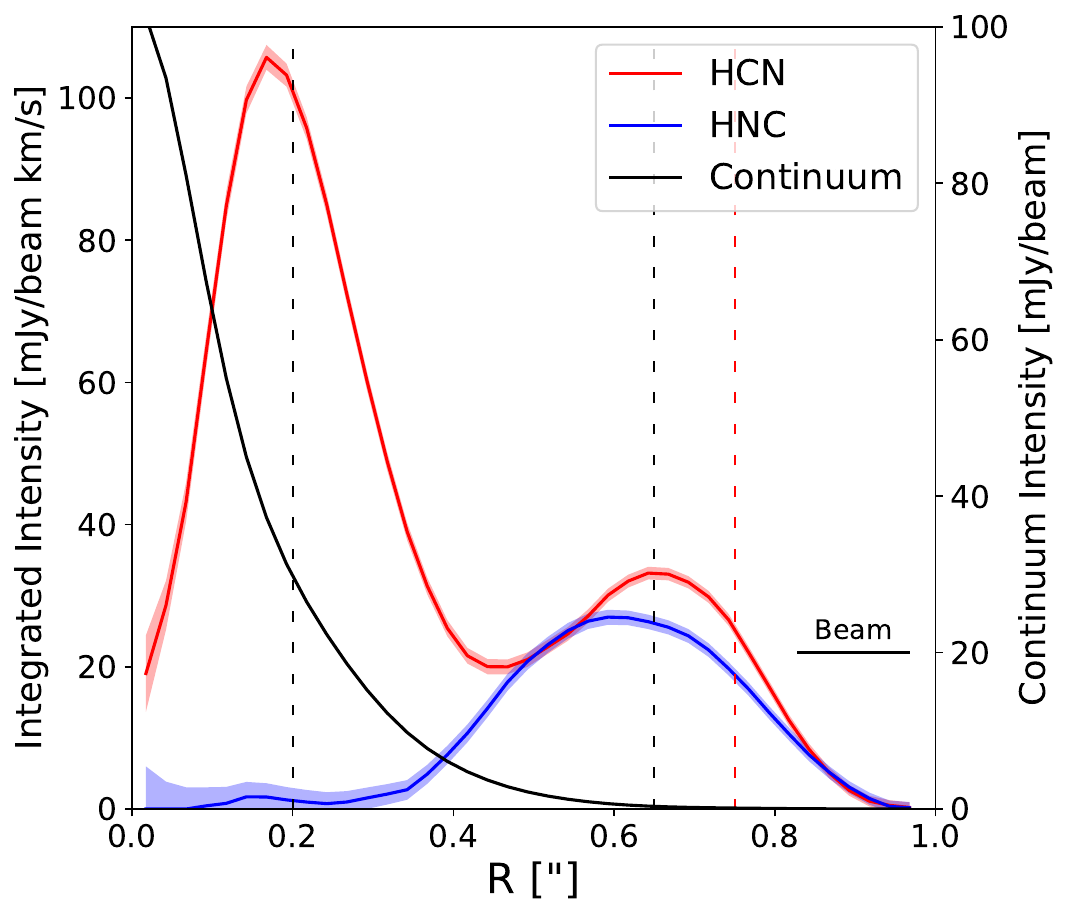}
    \caption{Radial emission distribution of HCN and HNC. The ribbons in light colors are 1$\sigma$ errors of the integrated intensity. The continuum distribution is plotted as a black solid line, with the scale on the right y-axis. The inner ring (0.2") and outer ring (0.65") are represented with black vertical dashed lines. The location of the HN$^{13}$C peak (at 0.75") is represented with red vertical dashed lines. The beam size ($\sim$0.13") is marked at the bottom right.}
    \label{fig:V883_radial}
\end{figure}

\begin{table}[t!]
\caption{Lines in focus.\label{tab:lines}}
\centering
\begin{tabular}{ccccc}
\toprule
Species & Frequency & $A_{ul}$ & n$_{\rm crit}$$^a$&$E_u$ \\
 & [GHz] & [s$^{-1}$] & [cm$^{-3}$]& [K] \\
\midrule
HCN $J$=3-2 & 265.886 & 8.36 (-4) & 9.6 (6) &25.52 \\
HNC $J$=3-2 & 271.981 & 9.34 (-4) & 6.2 (6) &26.11 \\
H$^{13}$CN $J$=3-2 & 259.011 & 7.73 (-4)& 8.9 (6)$^{\rm b}$& 24.86 \\
HN$^{13}$C $J$=3-2 & 261.263 & 8.28 (-4) & 5.5 (6)$^{\rm b}$& 25.08 \\
    
\bottomrule
\end{tabular}
\tabnote{
a(b) indicates a$\times$10$^{b}$. \par 
All data except for the critical density is from the CDMS database \citep{Muller2001}, queried with Splatalogue (https://splatalogue.online/). \\
$^{\rm a}$ The critical density is derived when the gas temperature is 25\,K.\\
$^{\rm b}$ The collision rates of H$^{13}$CN and HN$^{13}$C are assumed to be the same as those of HCN and HNC, respectively.
}
\end{table}

Figure \ref{fig:V883_radial} illustrates the azimuthally averaged radial distribution of the emission after correction for the inclination of 38.3° \citep{Cieza2016}. The distribution clearly shows the presence of ring structures. The continuum radial profile is relatively smooth and does not exhibit any rings. It can thus be concluded that the discrepancy between the continuum and HCN/HNC emission is most likely due to chemical reasons. The ring structure was previously predicted by \cite{Long2021}, but this is the first time it has been observed in a protoplanetary disk without sub-structures in the continuum emission. \par

In the inner ring, the lower limit of the HCN/HNC intensity ratio is approximately 10, based on the upper limit (3$\sigma$) of HNC intensity. In the outer ring, the intensity ratio is roughly equal to one ($\sim$1.2). 
However, the HCN and HNC lines may be optically thick in our observation, which limits the accuracy of the subsequent analysis. For this reason, we use their isotopologues, H$^{13}$CN J=3-2 (259.011\,GHz) and HN$^{13}$C J=3-2 (261.263\,GHz) lines. Although carbon isotope chemistry can lead to $^{13}$C fractionation, it is known that the abundance ratio of HCN/H$^{13}$CN and HNC/HN$^{13}$C are similar at the final stage of dense molecular clouds \citep{Colzi2020}. 
The bottom panels of Figure \ref{fig:V883} show the integrated intensity maps of H$^{13}$CN and HN$^{13}$C. As with HNC, HN$^{13}$C is undetected in the inner ring. It should be noted that some emission is visible in the vicinity of the center within the water sublimation region in the HN$^{13}$C integrated intensity map. This emission originates from CH$_2$DOH, rather than from HN$^{13}$C. Within the radius of water sublimation, COMs also sublimate and produce numerous lines blended with other lines (Jeong et al. submitted). The peak of HN$^{13}$C emission is on the southwest side of the disk, with distance of $\sim$0.75" from the center. The H$^{13}$CN/HN$^{13}$C intensity ratio in that point is $1.6 \pm 0.6$. Given the similarities in the properties of HCN and HNC transitions, as summarized in Table \ref{tab:lines}, we assume that the H$^{13}$CN/HN$^{13}$C line intensity ratio of $1.6 \pm 0.6$ represents the abundance ratio between HCN and HNC at the outer ring and seek to reproduce this ratio using our chemical model.

\section{Model}

In this study, we seek to construct a model that can explain the observed HCN/HNC ratio in the two rings of V883 Ori using a simple one-point chemical calculation. We do not intend to reproduce the 2D ring structures observed in the HCN and HNC emission or the absolute column densities of the molecules. This requires a detailed physical model of the V883 Ori disk to be combined with chemistry, which is beyond the scope of this study.

\subsection{Chemistry}
For the chemical calculation, we use a subroutine of PURE-C \citep{shlee2014, shlee2021,shlee2024}. PURE-C is an axisymmetric 2D thermochemical model that self-consistently calculates the gas and dust temperatures and the chemical abundances for a given density structure and the radiation field from the central star. The subroutine for chemical calculations can solve the chemical evolution in a specific physical condition of the density and temperature of gas and dust, cosmic ionization rate, and UV flux. 
The chemical network has been reduced from the original network used in \citet{shlee2021,shlee2024} by excluding chemical processes related to isotopes, such as those involving $^{13}$C and $^{15}$N. The network is sufficient for investigating the HCN and HNC abundance evolution \citep{Visser2018,shlee2021,shlee2024}.
The network considers the gas-phase chemistry, interactions between gas and grain surfaces, and grain surface chemistry. 
The initial abundances adopted in this study are presented in Table \ref{tab:init}. \par

For the barrier energies of reactions (\ref{reacH} and \ref{reacO}), we adopt the barrier energies of (200K, 20K) as fiducial values, following \cite{Hacar2020}. These values were able to reproduce the HCN/HNC abundance ratio observed from the Orion ISF, which is likely to have a similar star formation environment to V883 Ori. We also examine higher barrier energies to investigate the effect in Section \ref{sec:Barrier}.

\begin{table}[t!]
\caption{Initial abundances relative to total hydrogen atom used in the chemical model.\label{tab:init}}
\centering
\begin{tabular}{lc}
\toprule
Species & Abundance \\
\midrule
    H$_2$ & 5.00E-01\\
    He & 9.75E-02\\
    O & 1.80E-04\\
    N & 2.47E-05\\
    C$^+$ & 7.86E-05\\
    Mg$^+$ & 1.09E-08\\
    S$^+$ & 9.14E-08\\
    Si$^+$ & 9.74E-08\\
    Fe$^+$ & 2.74E-09\\
\bottomrule
\end{tabular}
\end{table}

\subsection{Physical conditions}

\begin{table*}[t!]
\caption{Physical parameters in the chemical model. \label{tab:params}}
\centering
\begin{tabular}{c|c|c}
\toprule
Parameters          & \multicolumn{2}{c}{Values}               \\
                    & Inner Ring         & Outer Ring          \\ \midrule
n  [cm$^{-3}$]      & $2.2\times 10^{10}$& $6.0\times 10^{7}$  \\
T (pre-burst) [K]   & 29                 & 10                  \\
T (after burst) [K] & 76                 & 25                  \\
UV [Draine Field]   & 0                  & 0                   \\
CR [s$^{-1}$] $\rm^a$& $10^{-18}$, $\mathbf{10^{-17}}$, $10^{-16}$ & $10^{-18}$, $\mathbf{10^{-17}}$, $10^{-16}$ \\
\bottomrule
\end{tabular}
\tabnote{$\rm ^a$ The values for the cosmic ray ionization rates of the fiducial models are written in boldface.}
\end{table*}

We assume that the observed HCN and HNC emission originates from the disk midplane. While previous modeling studies of protoplanetary disks have indicated that the observed HCN may originate from the warm molecular layer in the inner disk \citep{shlee2021, Long2021}, these studies primarily focus on disks in the quiescent accretion phase. In contrast, V883 Ori is in a burst accretion phase, which results in significantly elevated temperatures and a heated disk midplane. Consequently, the conditions derived from previous studies cannot be directly applied to V883 Ori. In V883 Ori, the HCN emission distribution in the inner region is similar to that of various COMs, whose emission region is close to the midplane. Therefore, we adopt the physical condition of the disk midplane in V883 Ori for our one-point chemical calculation of HCN and HNC. \par
The disk model of \cite{Andrews2009} was used for the density structure. The density in the midplane, $\rho(R, \Theta=\pi/2)$, is calculated with the following equations:
\begin{align*}
    \Sigma(R) = \Sigma_c \left(\frac{R}{R_c}\right)^{-\gamma} \exp\left[-\left(\frac{R}{R_c}\right)^{2-\gamma}\right], \\
    \rho(R,\Theta) = \frac{\Sigma(R)}{\sqrt{2\pi}Rh}\exp\left(-\frac{1}{2}\left(\frac{\pi/2 -\Theta}{h}\right)^2\right),\\ \hspace{1cm} h = h_c \left(\frac{R}{R_c}\right)^\psi,
\end{align*}
where $\Sigma$ is the gas surface density, and $\Sigma_c$ is the normalization factor of surface density. $\gamma$ is the power law index. $h$ is the angular scale height, which follows a power law with a flaring index $\psi$ and a characteristic scale height $h_c$ at $R_c$. The angular scale height is connected to the physical width of the Gaussian vertical density profile, $h_R = Rh$.
We adopt the model parameters of $R_c=$ 75\,au, $\Sigma_c=$\,35\,g\,cm$^{-2}$, $\gamma=$\,1, $h_c=$\,0.1, and $\psi$=\,0.25, from the model of V883 Ori by \cite{Leemker2021}.
According to the equations above, the densities for the inner and outer rings located at $R=$\,80\,au (0.2") and $R=$\,300\,au (0.75") are calculated as $2.2\times 10^{10}$\,cm$^{-3}$ and $6.0\times 10^7$\,cm$^{-3}$, respectively. \par
The current temperature can be calculated from the optically thick lines. We used the HCN line profile before continuum subtraction. The peak intensity of the optically thick HCN line is 48\,mJy/beam and 13\,mJy/beam for the inner ring and outer ring, respectively. The rms noise is 2.2\,mJy/beam. This corresponds to $76\pm3$\,K and $25\pm3$\,K, converted using the Planck function. We adopt these values as the temperature after the burst. \cite{Cieza2016} claimed the bolometric luminosity of the star before the burst could be $\sim$6\,L$_\odot$, assuming an age of 0.5\,Myr. When we adapt the bolometric luminosity in the current burst phase to be about 285\,L$_\odot$ \citep{Strom1993}, the temperature before the burst is scaled down with a factor of $(6/285)^{1/4}$. Therefore, we can assume the pre-burst temperature to be $29\pm2$\,K and $10\pm2$\,K, respectively. \par
The UV radiation and cosmic ray ionization rate (CR) are poorly constrained compared to density and temperature. We reasonably assume the UV radiation is fully shielded in the disk midplane, while CR is left as a free parameter. Furthermore, we assume that CR is uniform throughout the midplane. The physical parameters used in the chemical model are presented in Table \ref{tab:params}. Sections \ref{sec:UV} and \ref{sec:CR} discuss the impact of UV radiation and cosmic rays on the model results. \par
The model consists of two distinct phases, the quiescent phase and the burst phase, with respective timescales of 0.5 million years and 500 years. An age of 0.5 Myr is assumed for V883 Ori to estimate its pre-burst luminosity \citep{Cieza2016}. The typical timescale of an FU Ori-like outburst is on the order of centuries \citep{Connelley2018}. V883 Ori has been in an outburst phase since at least 1888 \citep{Strom1993}, though the exact duration remains unknown. In our model, we assume a long burst phase duration of 500 years. However, the real outburst period may be determined if the HCN/HNC ratio changes significantly over time. We assume that the accretion burst only affects the disk temperature, without altering the density structure.

\section{Results}
\subsection{Abundances}
\begin{figure*}
    \centering
    \includegraphics[width=\columnwidth]{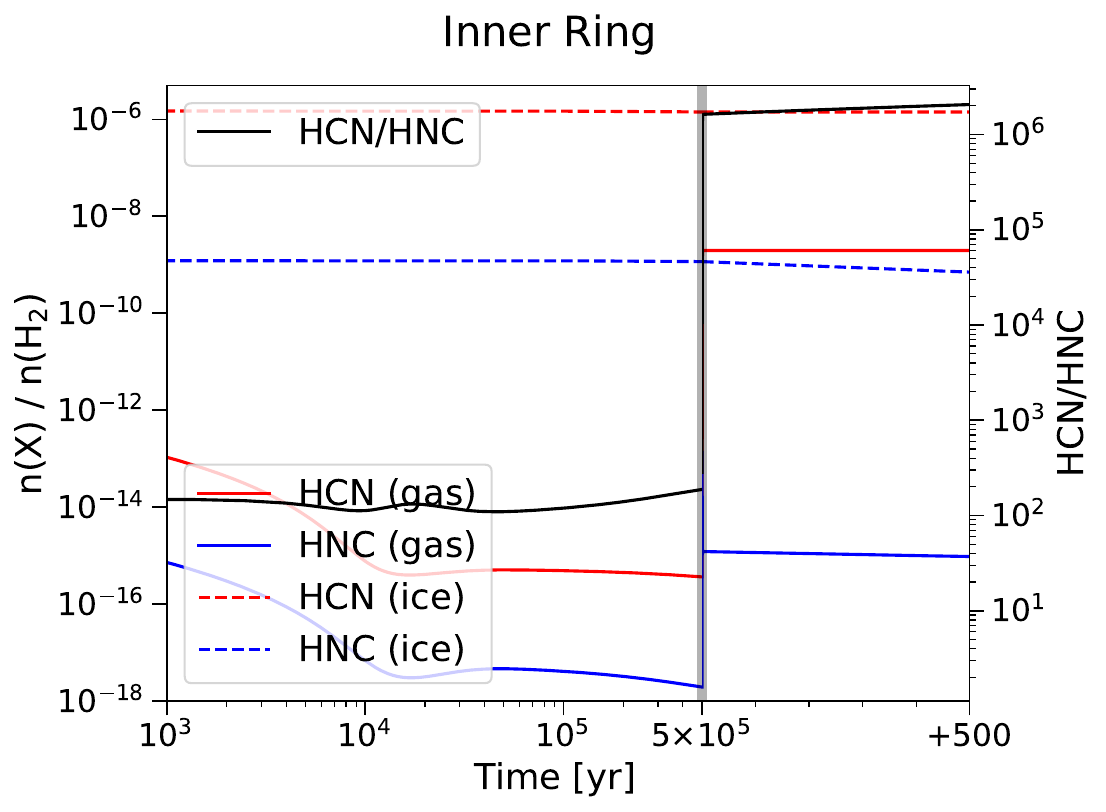}
    \includegraphics[width=\columnwidth]{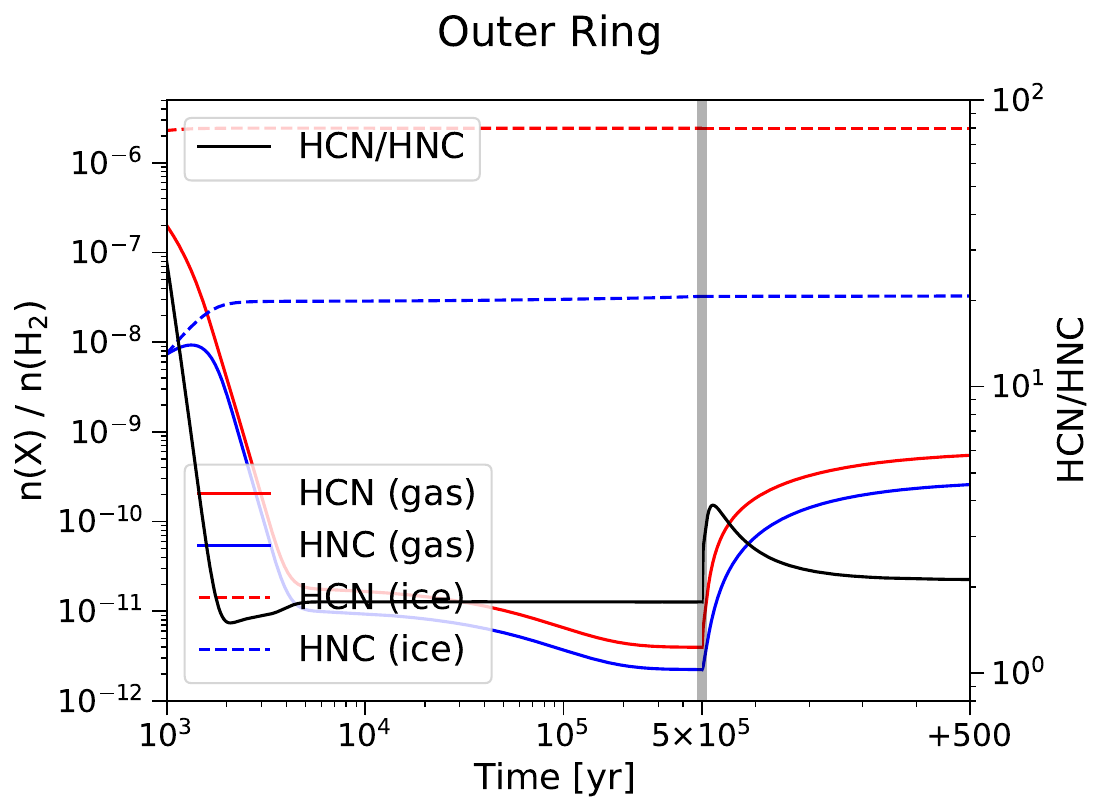}
    \caption{Chemical evolution for the inner ring (left) and the outer ring (right). Gas phase and ice phase abundances are shown with solid and dashed lines on the left y-axis. The HCN/HNC abundance ratio in the gas phase is depicted with a black solid line on the right y-axis. The vertical thick gray line represents the time of the burst ($5\times 10^5$ years).}
    \label{fig:V883_2points}
\end{figure*}

\begin{figure*}
    \centering
    \includegraphics[width=\columnwidth]{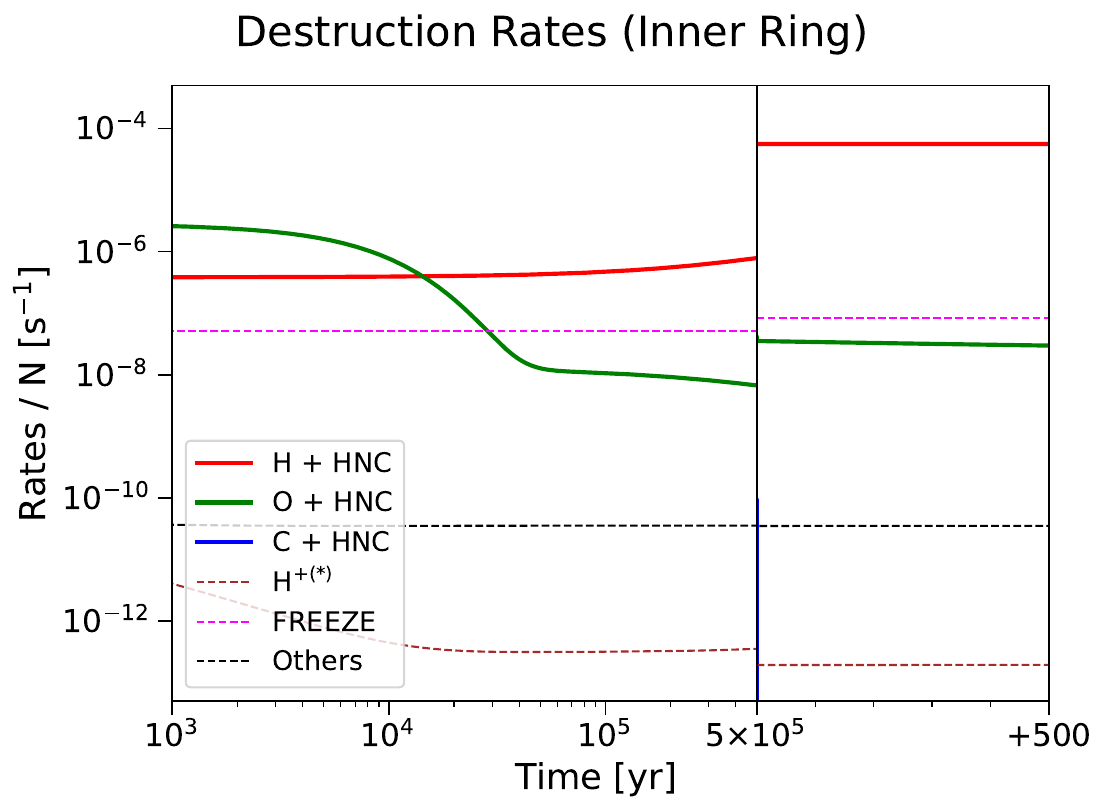}
    \includegraphics[width=\columnwidth]{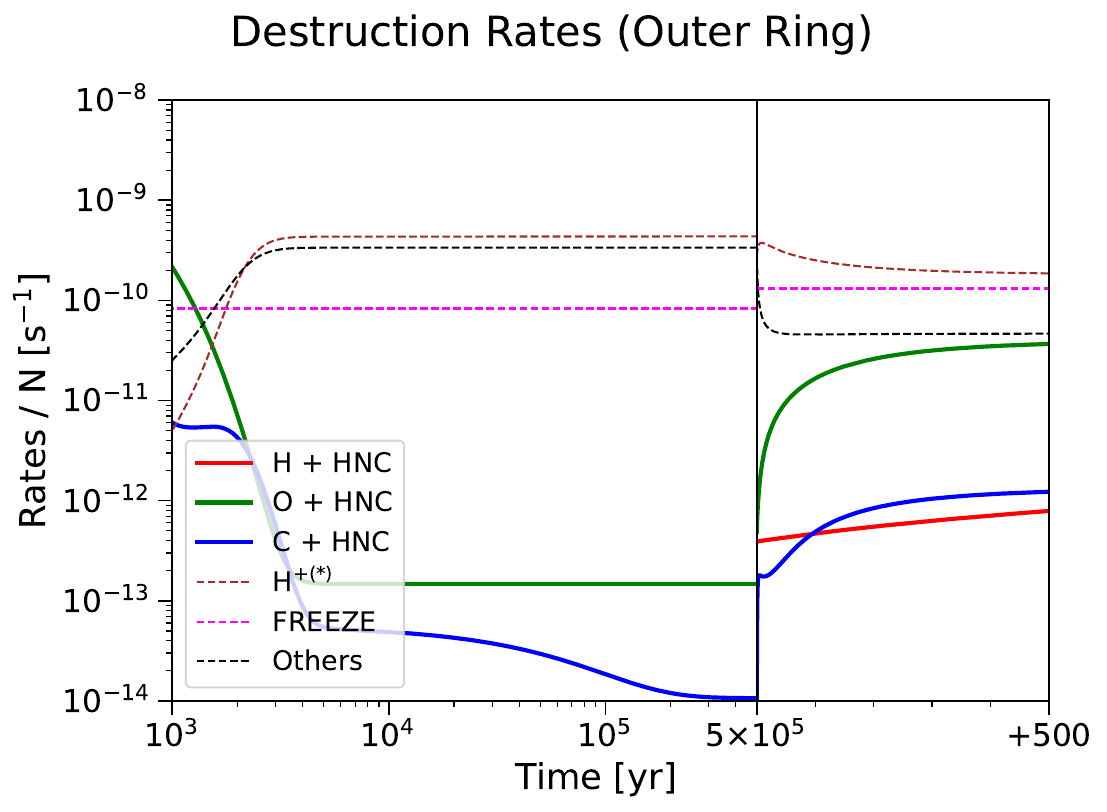}
    \caption{Destruction rates in the two rings. Reactions that happen to both HCN and HNC are drawn in dashed lines, while reactions only with HNC are drawn in solid lines. Note that H$^{+}$ reacts with both HCN and HNC with almost equal rates. However, it turns HNC into HCN and HCN into HCN$^{+}$. }
    \label{fig:V883_dest}
\end{figure*}

\begin{figure*}
    \centering
    \includegraphics[width=\columnwidth]{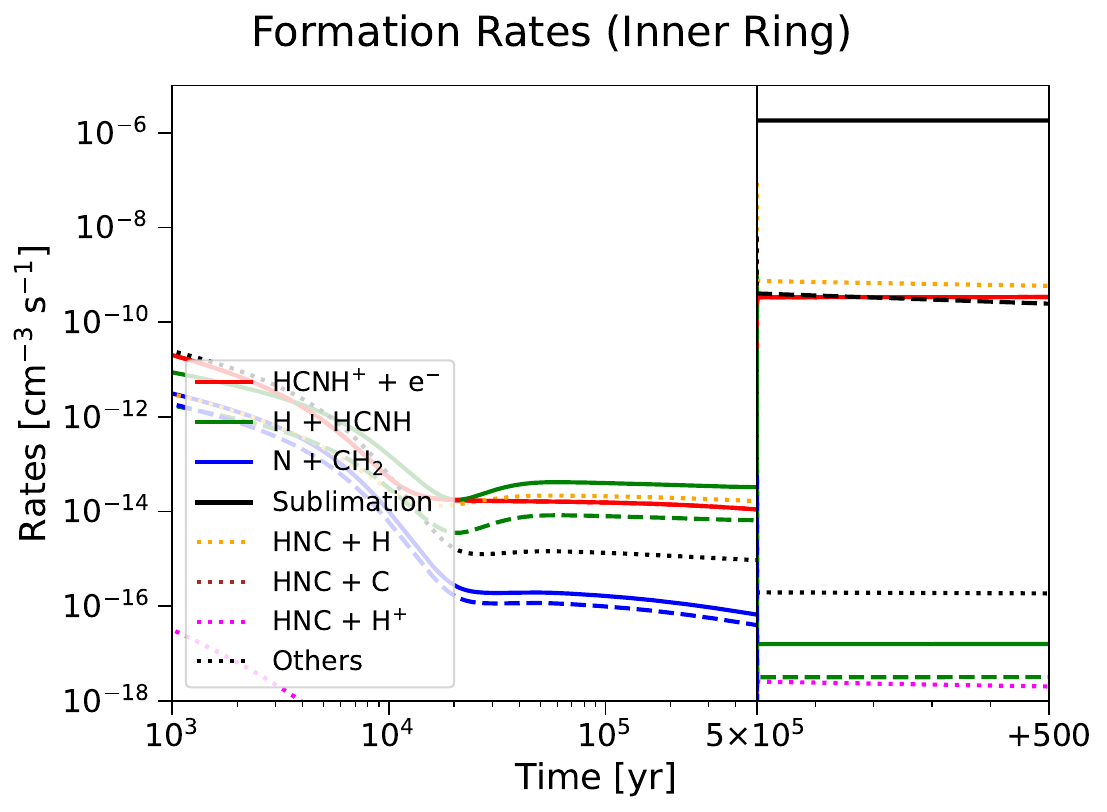}
    \includegraphics[width=\columnwidth]{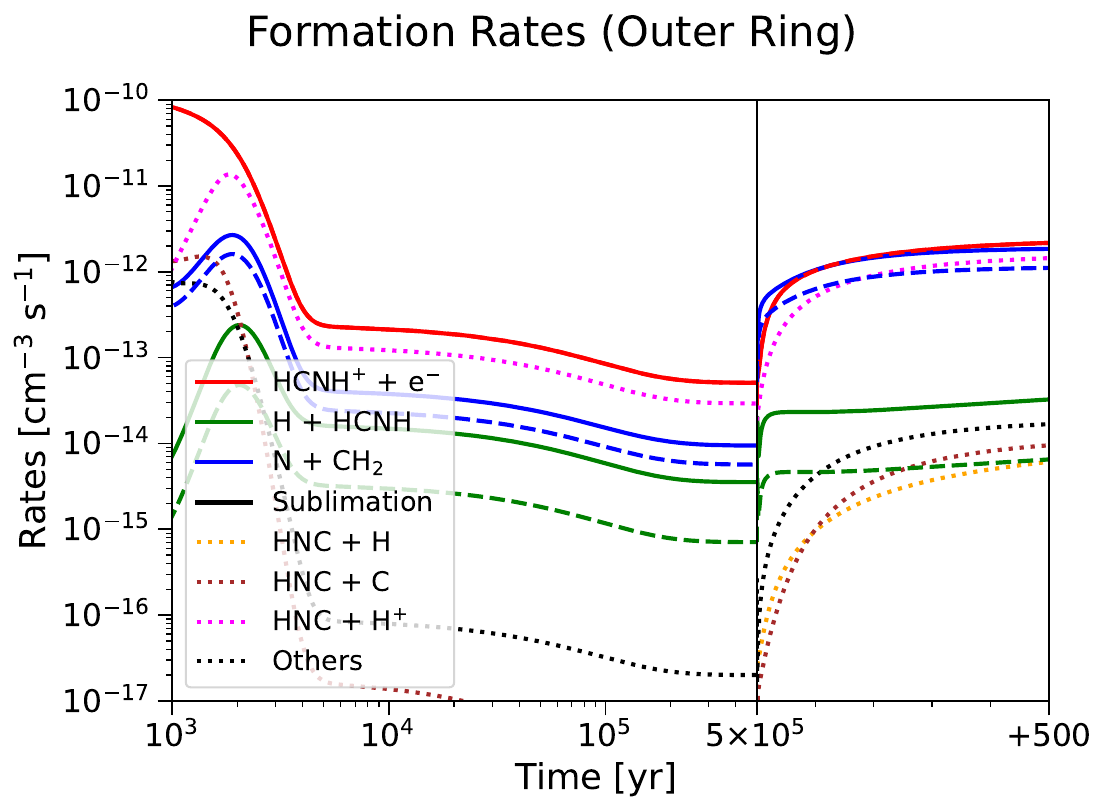}
    \caption{Formation rates in the two rings. Reactions that form both HCN and HNC are drawn in solid and dashed lines, respectively, since they have different formation rates. Reactions that only form HCN are drawn with dotted lines.}
    \label{fig:V883_form}
\end{figure*}

Figure \ref{fig:V883_2points} shows the best-fit result with CR of $10^{-17}$\,s$^{-1}$. In the initial quiescent phase, even in the inner ring, the majority of HCN and HNC are present in the ice phase, due to the significantly lower gas temperature (29\,K) compared to their sublimation temperature ($\sim$80\,K). Their gas phase abundances decrease over time, reaching equilibrium after $\sim$10$^4$ years, at $\sim$10$^{-15}$ and $\sim$10$^{-18}$, respectively. The ice phase abundances of HCN and HNC are approximately $10^{-6}$ and $10^{-9}$, respectively. \par

After the burst, the ice abundances remain relatively constant, whereas the gas abundances change significantly. The gaseous HCN abundance sharply increases immediately following the burst, reaching an equilibrium level of $\sim$10$^{-9}$. Additionally, the gas-phase abundance of HNC increases to $\sim$10$^{-15}$, arbeit at a much lower level than that of HCN. Consequently, the HCN/HNC abundance ratio during the burst is highly elevated, consistent with the observed ratio in the inner ring. \par
The outer ring has been modeled in the same way as the inner ring, with CR of $10^{-17}$\,s$^{-1}$. Before the burst, the abundance ratio reach equilibrium after $\sim$10$^4$ years. The gas phase abundances for HCN and HNC are both $\sim$10$^{-11}$, while the ice phase abundances are $\sim$2$\times$10$^{-6}$ and $\sim$ 5$\times$10$^{-8}$, respectively. 
Despite the temperature increase from 10\,K to 25 \,K during the burst, this cannot result in a notable change in the abundance ratio. The gas phase abundances of both molecules increase, yet the HCN/HNC abundance ratio remains at the low level of $\sim$2, within the error margin of the intensity ratio observed in H$^{13}$CN and HN$^{13}$C.

\subsection{Destruction and formation reactions}

Analyzing how the destruction and formation reactions of molecules vary over time enables an understanding of the mechanisms underlying the observed changes in abundance, as illustrated in Figure \ref{fig:V883_2points}. Figure \ref{fig:V883_dest} shows the destruction reaction rates of HCN and HNC in the two rings. The reaction rates of ions, including H$_{3}$O$^{+}$, HCO$^{+}$, C$^{+}$, and H$_{3}^{+}$, are identical for both HCN and HNC. The aforementioned reactions are summed up and illustrated by a black dashed line. Additionally, the freeze-out rate to the dust surface (depicted with a pink dashed line) is identical. However, the following reactions can change the HCN/HNC abundance ratio:
\begin{align}
    \mathrm{HNC + H} &\rightarrow \mathrm{HCN + H} \label{reacH} \tag{R1}\\
    \mathrm{HNC + O} &\rightarrow \mathrm{NH + CO} \label{reacO} \tag{R2}\\
    \mathrm{HNC + C} &\rightarrow \mathrm{HCN + C} \label{reacC} \tag{R3}\\
    \mathrm{HNC + H^{+}} &\rightarrow \mathrm{HCN + H^{+}} \label{reacHp1} \tag{R4}\\
    \mathrm{HCN + H^{+}} &\rightarrow \mathrm{HCN^{+} + H} \label{reacHp2} \tag{R5}
\end{align}

In the inner ring before the burst, the reaction of HNC with H (\ref{reacH}, red solid line) and O (\ref{reacO}, green solid line) converts HNC into HCN or destroys HNC. These reactions are further enhanced after the burst, increasing the HCN/HNC abundance ratio. Although the barrier energy of reaction \ref{reacH} (200\,K) is much higher than the gas temperature of 29\,K and 76\,K, the reaction can still occur if the abundance of the reactant is sufficiently large. It should be noted that the term "barrier energy" should be understood as a relative concept rather than as an absolute physical barrier. It is one of the coefficients in the modified Arrhenius equation:
\begin{equation*}
    k = \alpha \left(\frac{T}{300 \rm \,K}\right)^\beta e^{-E_{\rm b}/T},
\end{equation*}
where $E_{\rm b}$ is the energy barrier of the reaction. \par
In the outer ring before $\sim$10$^4$\,years, the reactions of HNC with O (\ref{reacO}, green solid line) and C (\ref{reacC}, blue solid line) are the most dominant destruction reactions. This is why the abundance ratio is not unity even during the early stages of the outer ring. As time passes, the abundance of atomic O and C decreases, and the reaction rates of \ref{reacO} and \ref{reacC} become negligible. Since the rate of HNC destruction or conversion to HCN decreases, the HCN/HNC abundance ratio returns back to a smaller value. After the burst, despite the temperature increase from 10\,K to 25\,K, the destruction rate remains relatively unchanged, resulting in no alteration to the abundance ratio. \par

Figure \ref{fig:V883_form} shows the formation reaction rates of HCN and HNC in the two rings. Despite the previous assumption that HCN and HNC are formed primarily through the dissociative recombination of HCNH$^{+}$ \citep[red solid line,][]{Mendes2012}, there are additional reactions that also contribute to their formation. The reactions of atomic H with HCNH (green lines) and atomic N with CH$_{2}$ (blue lines) can also form both HCN and HNC at a rate comparable to that of the dissociative recombination of HCNH$^{+}$. Other reactions, such as N + CH$_2$ and N + HCO, also result in the formation of HCN. The minor reactions that form HCN are depicted with black dotted lines. Additionally, the destructive reactions of HNC (\ref{reacH}, \ref{reacC}, \ref{reacHp1}, colored dotted lines) also result in the formation of HCN. \par

In the inner ring, before the burst, the dissociative recombination of HCNH$^{+}$ and the reaction of atomic H and HCNH are dominant. While the first reaction has a 1:1 branching ratio, the latter produces more HCN than HNC. This results in an increase in the HCN/HNC ratio in the gas phase, which is subsequently reflected in the ratio on grain surfaces. After the burst, the primary formation pathway for gaseous HCN and HNC is sublimation from the grain surface, as indicated by the black lines. This significantly increases the gas phase abundance ratio, since the abundance of HCN is much greater than HNC on the grain surface in the pre-burst phase. The HCN and HNC ice on the grain surface is formed by freeze-out from the gas phase. Hence, this ice phase abundance ratio originates from the gas phase abundance ratio at earlier times, when the conversion of HNC to HCN occurs immediately after the HNC formation by the reactions with the abundant H, O, and C atoms. During the burst, the gas phase abundance ratio is furthermore increased due to destructive reactions of HNC (\ref{reacH}, \ref{reacO}). \par
In the outer ring, the dominant formation mechanism is the dissociative recombination of HCNH$^{+}$, as was previously known. The reaction of HNC and H$^{+}$, which destroys HNC and creates HCN, is another dominant reaction for the formation of HCN, with a similar reaction rate to the dissociative recombination of HCNH$^{+}$. The change in reaction rates of temperature-sensitive reactions due to the burst is minor. \par

In conclusion, the HCN/HNC ratio within the inner ring is determined by the sublimation of HCN and HNC from the grain surface. This ratio is, therefore, dependent on the ice abundances that were developed during the pre-burst phase. During the burst, the reactions with atomic oxygen and hydrogen are enhanced in the inner ring due to the elevated temperature, converting HNC to HCN and thus further increasing the HCN/HNC ratio. In contrast, the temperature in the outer ring is considerably lower than the sublimation temperature of HCN and HNC, which hinders sublimation. Additionally, the reactions with atomic H, O, and C are relatively inactive due to the low temperature. The reaction rates and abundances of HCN and HNC remain unchanged before and after the burst. The slight difference in the reaction rates of formation mechanisms results in the HCN/HNC ratio of approximately 2.

\section{Discussion}

\subsection{Effect of barrier energies in reactions} \label{sec:Barrier}
In the models presented above, we used the low barrier energies of 200\,K and 20\,K for the reaction of HNC with atomic H (\ref{reacH}) and O (\ref{reacO}), respectively. These values were proposed by \cite{Hacar2020} in order to explain the abundance ratios observed in the Orion filament. However, quantum calculations propose much higher barrier energies \citep{Talbi1996, Graninger2014, Lin1992}. We tested how the calculated abundance ratio would change in the inner ring if the higher barrier energies were adopted. The barrier energies from KIDA \citep{Wakelam2012}, which are 2000\,K for \ref{reacH} and 1130\,K for \ref{reacO}, were tested.

\begin{figure}[t]
    \centering
    \includegraphics[width=\columnwidth]{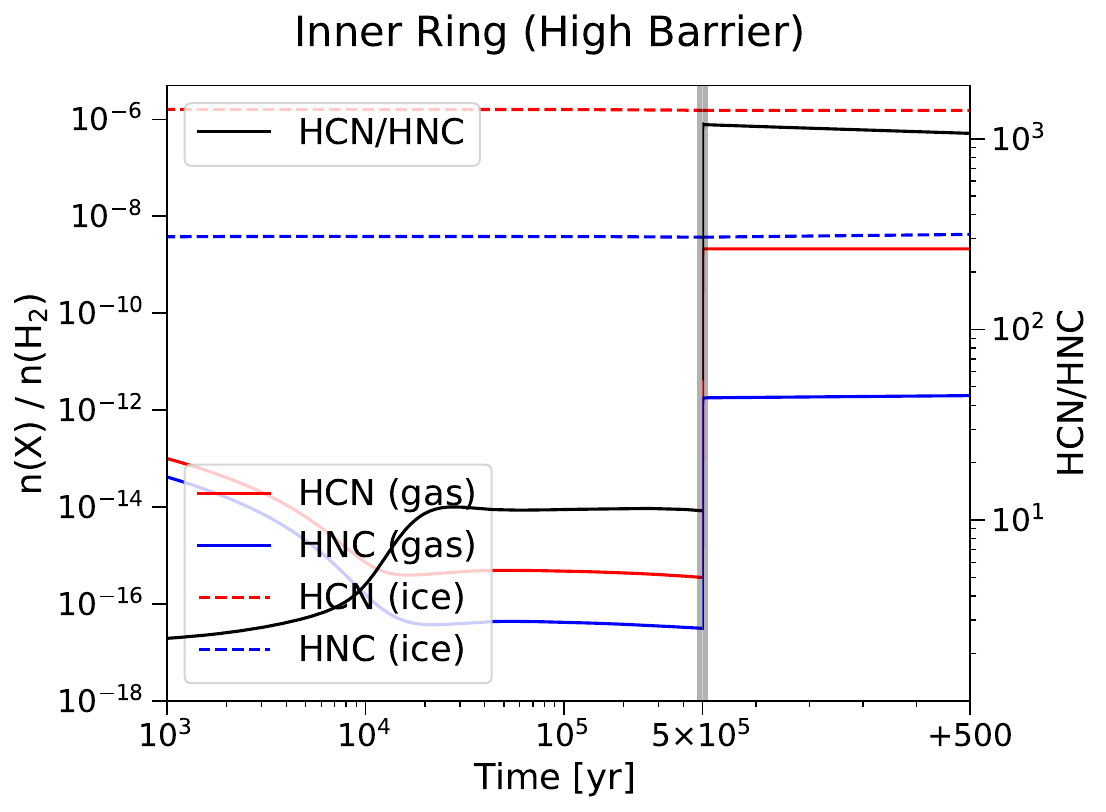}
    \caption{Alternate chemical model representing the inner ring with high barrier energies for the HNC destructive reactions.}
    \label{fig:V883_highE}
\end{figure}

\begin{table}[t!]
\caption{Abundances and densities of HNC in different simulations 200 years after burst. \label{tab:HNC}}
\centering
\begin{tabular}{c|c|c|c}
\toprule
Parameters           & \multicolumn{2}{c|}{Inner Ring}            & Outer Ring \\
                     & Low Barrier         & High Barrier         & \\
\midrule
n(HNC) / n(H$_2$)    & $1.1\times10^{-15}$ & $1.9\times 10^{-12}$ & $1.4\times10^{-10}$\\
n [cm$^{-3}$] & $2.2\times10^{10}$     & $2.2\times10^{10}$      & $6.0\times10^7$ \\
n(HNC) [cm$^{-3}$]   & $1.2\times10^{-5}$  & $2.1\times10^{-2}$   & $4.2\times10^{-3}$ \\
\bottomrule
\end{tabular}
\end{table}

Figure \ref{fig:V883_highE} shows the calculated abundances and the abundance ratio when high barrier energies were used.
Despite the lack of activation of reactions \ref{reacH} and \ref{reacO}, a high abundance ratio appears after the burst. This abundance ratio can be attributed to the ratio developed in the ice during the pre-burst phase. 
As expected, after the burst, the abundance of gaseous HNC is higher in the model with high barrier energies ($\sim$2$\times$10$^{-12}$) than in the model with low barrier energies ($\sim$10$^{-15}$). This is because the conversion reactions from HNC to HCN are not activated in the model with their high barrier energies. Therefore, if the barrier energies are as high as predicted from the theoretical calculations, the HNC line emission should be detected, which is not in accordance with our observational results. To determine the detection feasibility of the HNC line, we conducted a comparative analysis of the total molecular numbers between models. The number density of HNC is equal to the number density of H$_2$ multiplied by the abundance of HNC. Table \ref{tab:HNC} shows the HNC number density in the midplane for each model. If the energy barrier is indeed high, the HNC emission line should be observable in the inner ring since the HNC number density there is $\sim$5 times higher than that in the outer ring. The absence of HNC emission in the inner ring indicates that the barrier energies must be considerably lower than the theoretical values to account for the non-detection of HNC in this region.

\subsection{Effect of the UV radiation strength} \label{sec:UV}
\begin{figure}[t]
    \centering
    \includegraphics[width=\columnwidth]{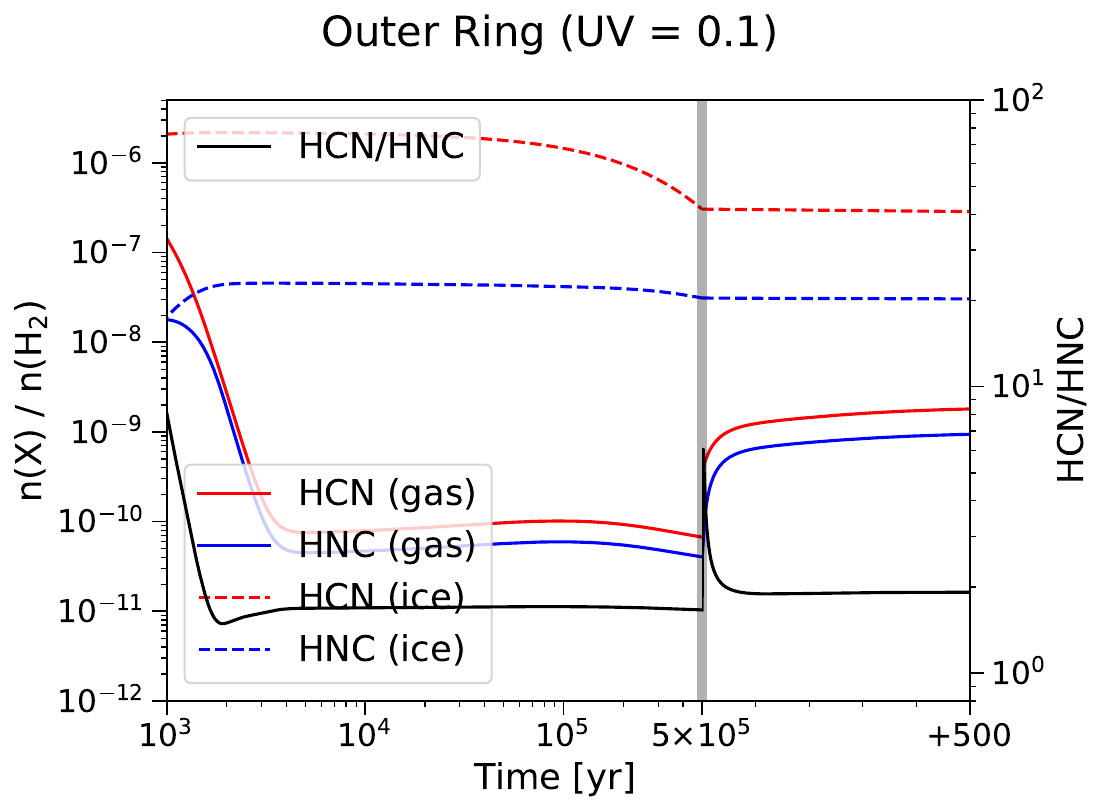}
    \caption{Alternate chemical model representing the outer ring with a UV field strength of 0.1 Draine field.}
    \label{fig:V883_UV}
\end{figure}

In the previous model calculations, we assumed the absence of a UV radiation field, as we adopted the physical conditions of the disk midplane. In the midplane at the inner ring, most UV photons from the central source could be shielded by dust grains due to the high surface density. However, in the outer ring, some UV may penetrate the midplane due to the relatively low surface density. During the burst phase, the UV radiation from the central source will also increase. Therefore, we investigated the impact of enhanced UV radiation in the outer ring on the abundance ratio. Self-shielding of CO, H$_{2}$, C, and N$_{2}$ is considered \citep[see,][]{shlee2021,shlee2024}. The column densities are calculated by multiplying the column density of total hydrogen molecules from the atmosphere to the midplane by their local abundances. \par
Figure \ref{fig:V883_UV} shows the abundances of HCN and HNC and their ratio at the outer ring when the dust attenuated UV flux is 0.1 Draine field after the burst. We used $4\times10^{-3}$ Draine field for the UV flux before the burst, given that the UV flux gets 25.2 times stronger when the bolometric luminosity increases from 6\,L$_\odot$ to 285\,L$_\odot$ \citep{shlee2015}. Normally UV radiation can photodissociate and ionize molecules to produce more C, O, H, and H$^{+}$. These neutral or ionic atoms can then react with HNC, thus reducing the abundance of  HNC and increasing the abundance ratio. However self-shielding protects the molecules in the midplane from undergoing dissociation, reducing the effect of UV radiation in chemistry. Therefore, the abundance ratio remained largely unaltered.

\subsection{Effect of the cosmic ray ionization rate} \label{sec:CR}

\begin{figure}[t]
    \centering
    \includegraphics[width=\columnwidth]{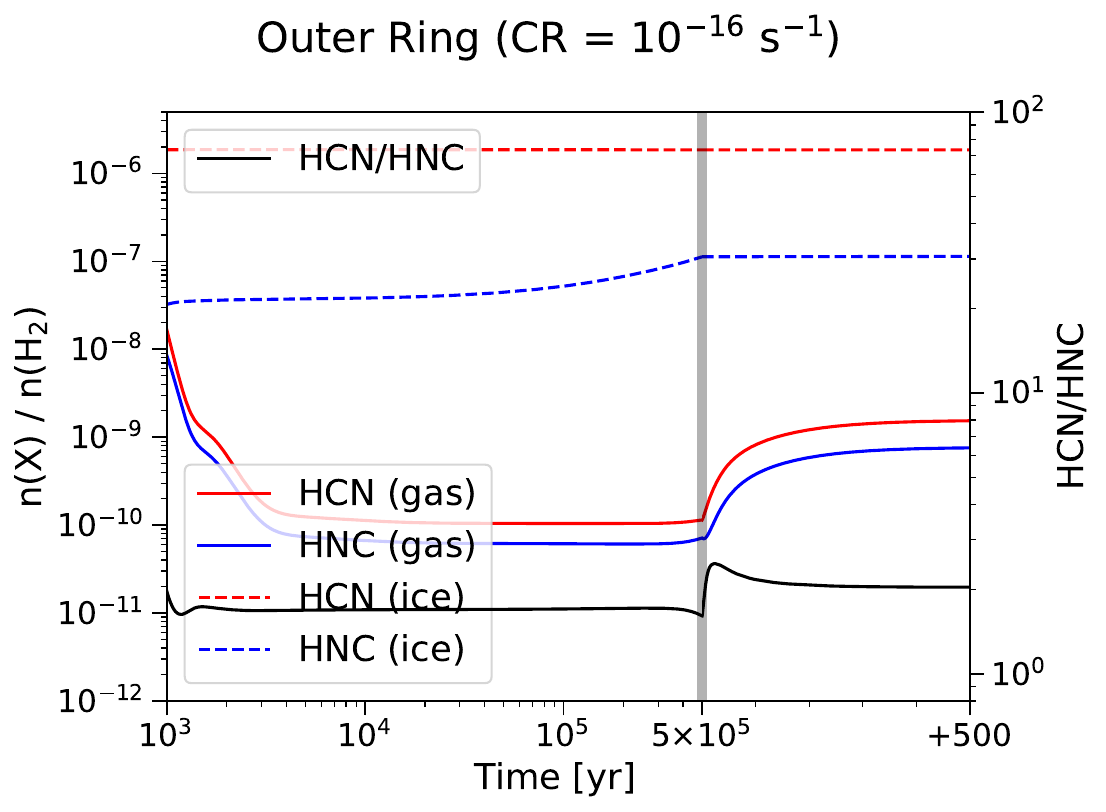}
    \includegraphics[width=\columnwidth]{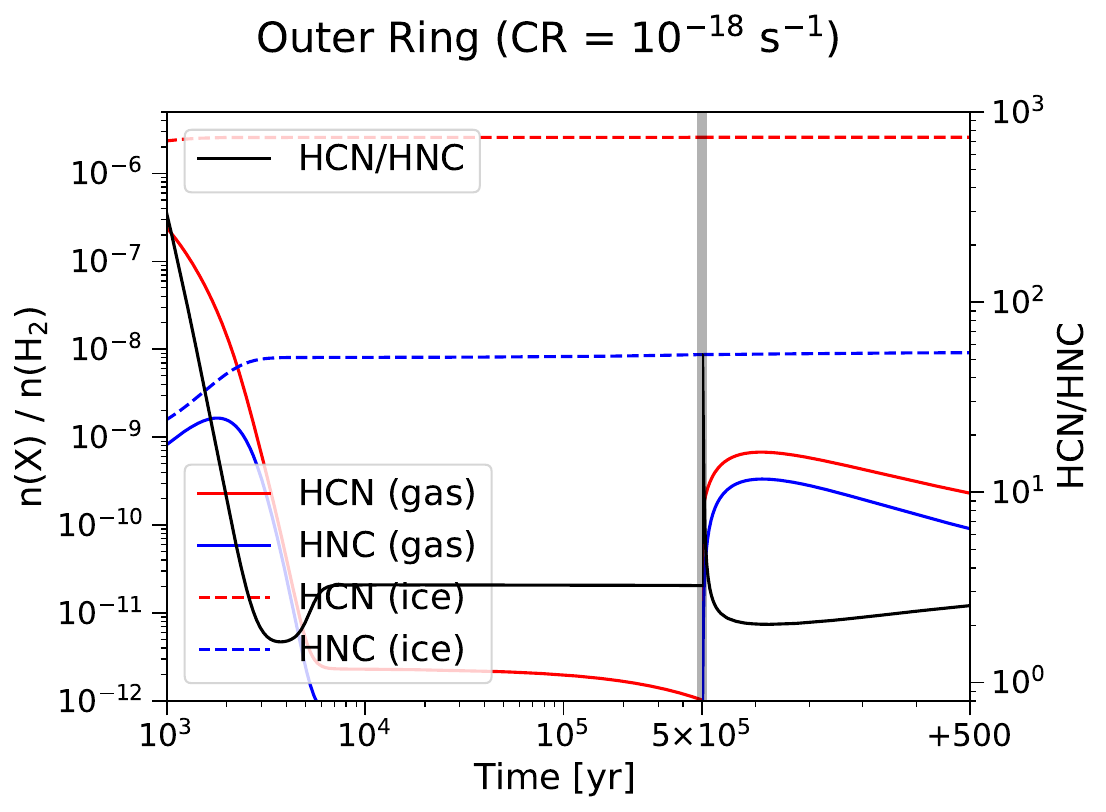}
    \caption{(Top) Alternate chemical model representing the outer ring with enhanced CR of $10^{-16}$\,s$^{-1}$. (Bottom) Same with the top panel, but with CR of $10^{-18}$\,s$^{-1}$.}
    \label{fig:V883_CR}
\end{figure}

We assumed that cosmic ray attenuation can be neglected and the inner and outer rings have the same CR value, given that the attenuation factor of cosmic rays is $\sim$100\,g\,cm$^{-2}$ \citep[the amount of matter that reduces the energy of the cosmic ray passing through it by a factor $e$;][]{Umebayashi1981,Fogel2011}. This value is much larger than the disk column density at the inner ring of V883 Ori \citep[$\sim$11\,g\,cm$^{-2}$;][]{Leemker2021}. The typical value used for the cosmic ray ionization rate is $10^{-17}$\,s$^{-1}$ \citep{Spitzer1968}. \par

We examined higher and lower CRs in the outer ring; Figure \ref{fig:V883_CR} illustrates the results for a CR of $10^{-16}$\,s$^{-1}$ (top) and $10^{-18}$\,s$^{-1}$ (bottom). After the burst, the HCN/HNC abundance ratio is approximately 2, consistent with the result of the fiducial model with CR = $10^{-17}$\,s$^{-1}$, in both cases. However, the absolute abundances of HCN and HNC are higher and lower by an order of magnitude, respectively, for the higher and lower CR values, before the burst.
To determine the cosmic ray ionization rate more accurately, it is necessary to consider the abundances of other molecular species.\par
It should be noted that CR can differ in different situations. This is demonstrated by models that consider the deflection of cosmic rays by magnetic fields \citep{Fuiji2022} and observations \citep{Aikawa2021}. In this study, we adopt the typical value of CR =\,$10^{-17}$\,s$^{-1}$ for both rings.

\section{Conclusion}
Molecular rings of HCN and HNC were observed in V883 Ori, with a very high HCN/HNC abundance ratio ($>$ 10) in the inner ring (R$\sim$80\,au) and a low ratio of $1.6 \pm 0.6$ at the outer ring (R$\sim$300\,au). In this paper, we explored the distinct chemical distribution of HCN and HNC using a chemical model.
\begin{enumerate}
    \item In the inner ring, the temperature likely increased to 76\,K (from 29\,K) in the burst phase, as derived from the peak intensity of optically thick molecular lines. The ice abundance ratio between HCN and HNC is intrinsically large, resulting in a high abundance ratio in gas after sublimation. However, the gas phase reactions, such as the reactions of HNC with atomic H and O convert HNC to HCN, further increasing the HCN/HNC ratio.
    \item In the outer ring, the temperature might have increased to 25\,K (from 10\,K), which was insufficient to significantly enhance sublimation or reaction rates for conversion. As a result, the abundance ratio remained consistent with that in the pre-burst phase. 
    \item The abundance ratios are likely less sensitive to the UV radiation strength and the cosmic ray ionization rate than density or temperature. 
\end{enumerate}

%%% ACKNOWLEDGMENTS (IF ANY) %%%%%%%%%%%%%%%%%%%%%%%%%%%%%%%%%%%%%%%%

\acknowledgments
This work was supported by the New Faculty Startup Fund from Seoul National University.
This work was supported by the National Research Foundation of Korea (NRF) grant funded by the Korea government (MSIT) (grant numbers 2021R1A2C1011718 and RS-2024-00416859).
This paper makes use of the following ALMA data: ADS/JAO.ALMA\#2019.1.00377.S.
ALMA is a partnership of ESO (representing its member states), NSF (USA), and NINS (Japan), together with NRC (Canada), NSC and ASIAA (Taiwan), and KASI (Republic of Korea), in cooperation with the Republic of Chile. The Joint ALMA Observatory is operated by ESO, AUI/NRAO and NAOJ.

%%% APPENDICES (IF ANY) %%%%%%%%%%%%%%%%%%%%%%%%%%%%%%%%%%%%%%%%%%%%%

%\appendix
%\section{Additional Figures}

%%% CALL LIST OF REFERENCES (natbib STYLE) %%%%%%%%%%%%%%%%%%%%%%%%%%
\bibliography{main}

\end{document}